\documentclass[conference]{IEEEtran}
\pdfoutput=1
\IEEEoverridecommandlockouts

\usepackage{cite}
\usepackage{amsmath,amssymb,amsfonts}
\usepackage{algorithmic}
\usepackage{graphicx}
\usepackage{textcomp}
\usepackage{xcolor}
\usepackage{hyperref}
\usepackage{tabularx}
\usepackage{booktabs}
\usepackage{multirow}
\usepackage{comment}

\def\BibTeX{{\rm B\kern-.05em{\sc i\kern-.025em b}\kern-.08em
    T\kern-.1667em\lower.7ex\hbox{E}\kern-.125emX}}
\begin{document}

\title{LLM-KT: A Versatile Framework for Knowledge Transfer from Large Language Models to Collaborative Filtering\\
\thanks{The work was partially prepared within the framework of the Basic Research Program at the National Research University Higher School of Economics (HSE).}
}

\author{
    \IEEEauthorblockN{Nikita Severin}
    \IEEEauthorblockA{
        HSE University\\
        nseverin@hse.ru
    }
    \and
    \IEEEauthorblockN{Aleksei Ziablitsev}
    \IEEEauthorblockA{
        MIPT 
    }
    \and
    \IEEEauthorblockN{Yulia Savelyeva}
    \IEEEauthorblockA{
        MIPT
    }
    \and
    \IEEEauthorblockN{Valeriy Tashchilin}
    \IEEEauthorblockA{
        Ural Federal University
    }
    \and 
    \IEEEauthorblockN{Ivan Bulychev}
    \IEEEauthorblockA{
        HSE University
    }
    \and
    \IEEEauthorblockN{Mikhail Yushkov}
    \IEEEauthorblockA{
        HSE University
    }
    \and
    \IEEEauthorblockN{Artem Kushneruk}
    \IEEEauthorblockA{
        HSE University
    }
    \and
    \IEEEauthorblockN{Amaliya Zaryvnykh}
    \IEEEauthorblockA{
        HSE University
    }
    \and
    \IEEEauthorblockN{Dmitrii Kiselev}
    \IEEEauthorblockA{
        HSE University
    }
    \and
    \IEEEauthorblockN{Andrey Savchenko}
    \IEEEauthorblockA{\textit{Sber AI Lab, HSE University}\\
        avsavchenko@hse.ru
    }
    \and
    \IEEEauthorblockN{Ilya Makarov}
    \IEEEauthorblockA{
        \textit{AIRI, ISP RAS}\\
        iamakarov@hse.ru
    }
}

\maketitle

\begin{abstract}




We present LLM-KT, a flexible framework designed to enhance collaborative filtering (CF) models by seamlessly integrating LLM (Large Language Model)-generated features. Unlike existing methods that rely on passing LLM-generated features as direct inputs, our framework injects these features into an intermediate layer of any CF model, allowing the model to reconstruct and leverage the embeddings internally. This model-agnostic approach works with a wide range of CF models without requiring architectural changes, making it adaptable to various recommendation scenarios.

Our framework is built for easy integration and modification, providing researchers and developers with a powerful tool for extending CF model capabilities through efficient knowledge transfer. We demonstrate its effectiveness through experiments on the MovieLens and Amazon datasets, where it consistently improves baseline CF models. Experimental studies showed that LLM-KT is competitive with the state-of-the-art methods in context-aware settings but can be applied to a broader range of CF models than current approaches.


\end{abstract}

\begin{IEEEkeywords}
Large Language Model (LLM), recommender systems, knowledge transfer, RecBole framework
\end{IEEEkeywords}

\section{Introduction}
Many recommender systems use Collaborative Filtering (CF) methods to model user preferences and match items to them~\cite{koren2021advances,shevchenko2024variability,kiselev2022exploration}. However, these models often struggle to understand nuanced relationships and adapt to dynamic user-item interactions~\cite{kumar2019predicting,severin2023ti}. To tackle this issue, applying Large Language Models (LLMs) for recommendations has been actively studied since LLMs offer new ways to represent knowledge with their strong reasoning capabilities. 



As a result, current studies have integrated LLMs into various stages of recommender systems, from open-world knowledge generation \cite{wang2023enhancing,wu2023leveraging} to candidate ranking \cite{wang2023recmind,sun2024large}. Since LLMs are expensive to use, recently, several works proposed to directly use LLM for improving the quality of CF models by performing knowledge transfer (e.g., KAR \cite{xi2023towards}, LLM-CF \cite{sun2024large}). They create textual features from reasoning chains of LLM and integrate them as input to CF models. However, such an approach limits their applicability to only context-aware models, making their direct usage impossible for other types of CF models that don't handle input features.


Given these limitations, we developed a method that extends the applicability of knowledge transfer from LLMs to a broader range of CF models. We introduce ``LLM-KT'', a novel framework that facilitates seamless integration with various CF models and provides a robust environment for testing and modifying the approach. Our framework enables efficient knowledge transfer by embedding LLM-generated features into the intermediate layers of CF models, training the models to reconstruct these features as a pretext task internally. This process allows the CF model to develop a more refined understanding of user preferences, resulting in more accurate recommendations. Experiments on two well-known benchmarks demonstrate that the proposed method significantly improves the performance of CF models (+ up to 21\% improvement in NDCG@10) while applying to a broader range of models than existing approaches and achieving results comparable to the state-of-the-art KAR \cite{xi2023towards} in context-aware setting.

\section{Proposed Method}
The primary concept of our knowledge transfer method is to let the CF model reconstruct user preferences from the LLM within a specific internal layer without altering its architecture. 
This approach mirrors the intuitive process of identifying user interests in the early layers and making recommendations based on these learned interests in the later layers.

\subsection{Proposed Knowledge Transfer}
Our method consists of the following steps.


\textbf{Profile Generation}. 
First, we use an LLM to generate short preference descriptions for each user based on their user-item interaction data. Following the terminology from \cite{xi2023towards, shu2023rah, zhang2023agentcf}, we refer to these descriptions as ``profiles''. Notably, any LLM-based framework can be used for this process, making our method flexible and adaptable to various scenarios \cite{xi2023towards,shu2023rah}. This flexibility allows the framework to accommodate different LLMs and approaches for generating personalized profiles, enhancing its adaptability to various use cases.



To maintain efficiency and reduce the number of calls to pretrained LLMs, we create these profiles by independently processing each user's interactions using customized interest reasoning prompts. For our dataset, we used the following prompt structure: ``\textit{Based on the user's ratings, provide a general summary of their preferences, paying attention to... The response should be organized into several parts...}''. As can be seen, we explicitly define the required components of the response to ensure the consistency of representations across users. A typical profile might look like this: ``\textit{It seems that you enjoy a mix of classic and modern movies, with a preference for...}''.

\textbf{Profile Embedding}. 
We apply a pre-trained text embedding model to convert the textual profiles into dense embeddings. In our experiments, ``text-embedding-ada-002'' is used.

\textbf{Training with auxiliary pretext task}. 
We add an auxiliary pretext task for a given CF model to reconstruct user profiles in a predefined internal layer. This is done without altering the model's architecture using a weighted sum of the model's loss and a reconstruction loss with the weight $\alpha \in [0,1]$:
\begin{equation}
 \mathcal{L}_{combined} = \alpha \cdot \mathcal{L}_{KT} + (1 - \alpha) \cdot \mathcal{L}_{model}
 \label{eq:Combined_loss}
\end{equation}

Here, $\mathcal{L}_{model}$ denotes the model-specific loss of a chosen CF model, e.g., BCE (Binary Cross Entropy) for interaction prediction, MSE (Mean Squared Error) for rating prediction, etc.). The reconstruction loss, denoted as $\mathcal{L}_{KT}$, is defined as follows. Let $P_u$ represent the profile embedding of user $u$, and let $Z_u$ denote the output of the $K_{th}$ layer of the CF model after processing the interactions of user $u$. Generally, the knowledge-transfer loss is defined as:
\begin{equation}
 \mathcal{L}_{KT}(Z_u, P_u) = \mathcal{L}_{reconstruct}(Z_u, Trans(P_u)),
 \label{eq:Loss_kt}
\end{equation}
where $Trans$ is a transformation function that aligns profile embedding to layer representation space.


For simplicity, we utilized a nonlearnable $Trans$ function to match the dimensions of profile embeddings with the dimensions of the model's internal layers. Although our framework supports any dimensionality reduction method, we selected UMAP \cite{mcinnes2018umap} for our experiments because it preserves the distances between embeddings more effectively than conventional PCA (Principal Component Analysis) \cite{mackiewicz1993principal} and can reduce dimensions to any desired number, unlike t-SNE (t-distributed Stochastic Neighbor Embedding)~\cite{van2008visualizing}. It enables the transformed embeddings to maintain relationships captured by LLM profiles. Our framework supports several options for reconstruction loss. AS RMSE (Root Mean Squared Error) produced the best results, we used it in the remaining part.

\subsection{Training process}

We train a CF model with LLM knowledge transfer for $N$ epochs during two phases:

\textbf{Phase 1: Knowledge transfer}. 
During the first $N/2$ epochs, we train the model using an auxiliary pretext task and a combined loss function, as defined in equation 1. This phase optimizes the model for learning to reconstruct LLM-generated features together with the recommendation task.

\textbf{Phase 2: Fine-tuning}. 
After completing the knowledge transfer, we remove the reconstruction loss and train the model for the remaining $N/2$ epochs, focusing solely on the prediction task to optimize the model for accurate recommendations.

\section{LLM-KT Framework}

\subsection{General Pipeline}
\begin{figure}[htbp]
\includegraphics[width=0.99\columnwidth]{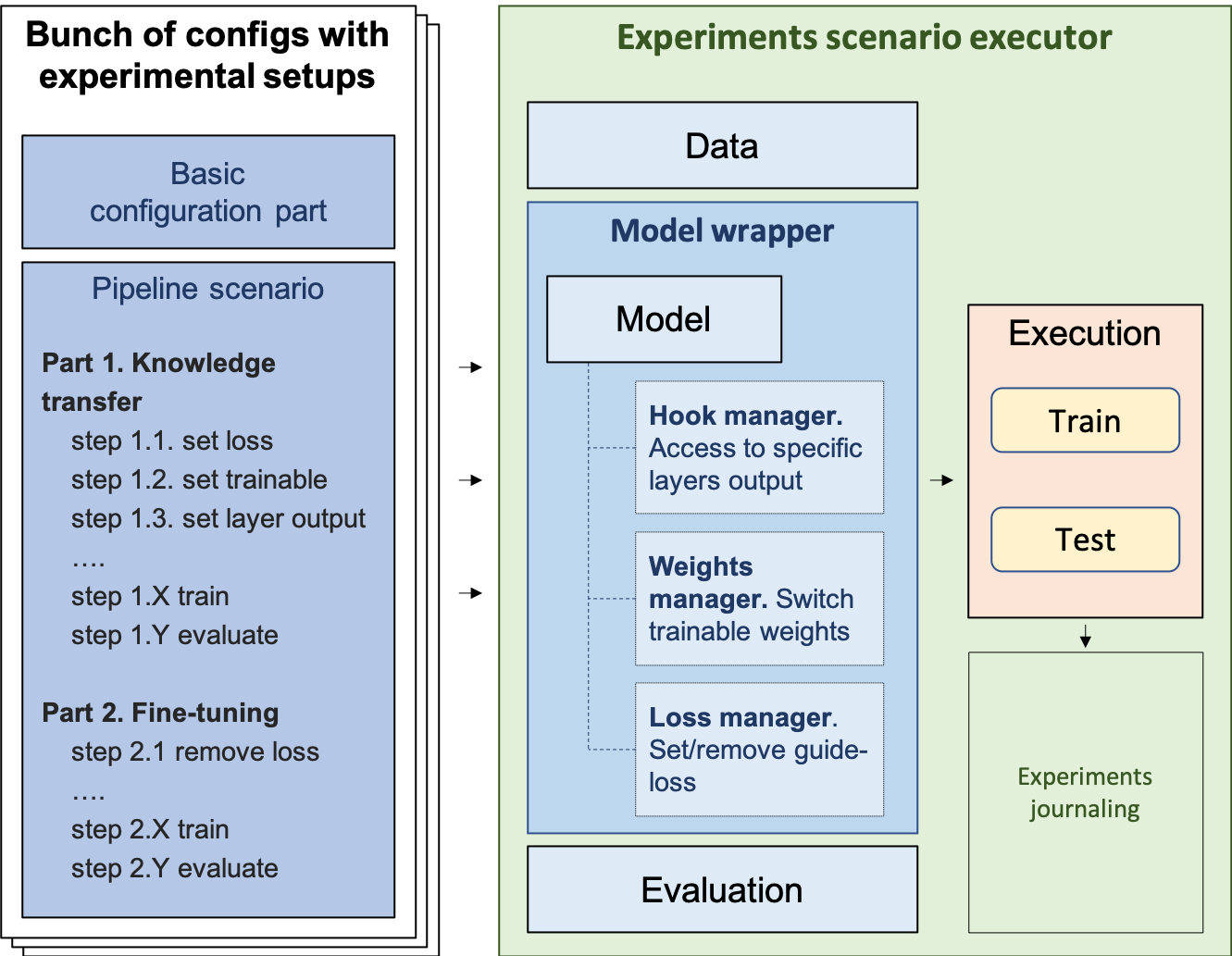}
  \caption{Proposed LLM-KT framework. A user can set up an experimental config, selecting the components and specifying the whole pipeline by declaring the sequence of predefined commands that will be applied. The framework supports various of them, including setting loss functions at different stages, selecting the layer number to conduct knowledge transfer into, freezing weights, and more.}
  \label{fig:pipeline}
\end{figure}

We developed a flexible experimentation framework (see Fig.~\ref{fig:pipeline}) on top of RecBole \cite{zhao2021recbole}, which allows seamless integration of LLM-generated features into CF models. The framework is designed to enable users to define complex experimental pipelines using a single configuration file, offering a versatile solution for knowledge transfer and finetuning processes in CF models. By supporting a variety of configuration options and predefined commands, it empowers researchers and developers to conduct experiments that explore different aspects of integration. 

\subsection{Framework Features}
\begin{itemize}

\item \textbf{Support for Any LLM-Generated Profiles}: The framework seamlessly integrates LLM-generated user profiles, supporting outputs from any existing methodology. This allows us to experiment with and compare different methods for profile construction.

\item \textbf{Flexible Experiment Configuration}: A key feature of the framework is its highly flexible configuration system for defining experimental pipelines. These configurations typically include standard RecBole setups (e.g., dataset splits for training, validation, and testing) and custom pipeline instructions. Users can define entire experiments by specifying sequences of predefined commands executed in order. Available commands include setting loss functions at various stages, selecting specific layers for knowledge transfer, freezing weights at chosen layers, and selecting subsets from the training dataset to transfer knowledge and finetune the CF model.

\item \textbf{Batch Experiment Execution and Comparison}: The framework enables users to run multiple experiments in batches, facilitating a more efficient and streamlined experimentation workflow.

\item \textbf{Analytical Tools}: Following the execution of experiments, the framework provides built-in tools for result analysis, allowing users to compare outcomes through visualizations and other analytical methods.
\end{itemize}

\subsection{Internal Structure}
The core of the framework is the Model Wrapper, which acts as an interface between the configuration and the underlying CF model. This wrapper manages key aspects of model manipulation through specialized components:

\begin{itemize}
  \item \textbf{The Hook Manager} provides access to the outputs of specific layers within the model, enabling detailed analysis and extraction of intermediate representations.
  \item \textbf{The Weights Manager} controls the freezing and unfreezing of trainable parameters, making it easy to apply selective finetuning strategies.
  \item \textbf{The Loss Manager} facilitates adding or removing custom losses, supporting advanced experimentation with different loss functions across various training stages.
\end{itemize}

The framework also includes an execution module for the training and testing phases and a journaling system that logs experimental outcomes for subsequent evaluation.

\section{Experimental Setup}



\textbf{Scenarios Under Analysis}. We analyzed our method from two perspectives. First, we evaluated its applicability to traditional CF models that rely only on user-item interaction data, showing that our framework enhances these models by capturing nuanced relationships through LLM-generated profile reconstruction. Second, we examined its use with context-aware models, where current LLM knowledge transfer methods pass LLM-generated features as input.

\textbf{Datasets}. We conducted experiments on two conventional datasets (Table~\ref{tab:datasets}), namely, \href{https://aws.amazon.com/datasets/}{Amazon ``CD and Vinyl''} (CDs) and \href{https://grouplens.org/datasets/movielens/}{MovieLens} (ML-1M). In all our experiments, we split the dataset into time-ordered training, validation, and test sets with ratios of 70-10- 20\%, following previous studies \cite{covington2016deep,ji2023critical}.

\begin{table}[h]
\centering
\small
\caption{Dataset statistics}
\label{tab:datasets}
\begin{tabularx}{\columnwidth}{XXXXX}
\toprule
\textbf{Dataset} & \textbf{Users} & \textbf{Items} & \textbf{Interactions} & \textbf{Sparsity($\%$)}\\
\midrule
\textbf{ML-1M} & 6,041 & 3,707 & 1,000,209 & 95.53 \\
\midrule
\textbf{CDs} & 4,558 & 7,784 & 194,242 & 99.45 \\
\bottomrule
\end{tabularx}
\end{table}

\textbf{Baselines:} To test the effectiveness of our approach, we selected three widely used neural CF baselines, each representing different architecture types:

\begin{itemize}
\item \textit{NeuMF} \cite{he2017neural}: Neural matrix factorization.
\item \textit{SimpleX} \cite{mao2021simplex}: An efficient CF model using contrastive learning.
\item \textit{MultVAE} \cite{liang2018variational}: A model based on the Variational Autoencoder (VAE).
\end{itemize}

In the context-aware scenario, where models can leverage input features, we compared our approach to the state-of-the-art knowledge transfer framework, KAR. The following baselines were selected for analysis:

\begin{itemize}
\item \textit{DCN} \cite{wang2017deep}: A cross network model.
\item \textit{DeepFM} \cite{guo2017deepfm}: A neural model based on factorization machines.
\end{itemize}

In all experiments, we ran at least $N=70$ epochs of each baseline to ensure the absence of the grokking effect. 

\section{Experimental Results}
\begin{table*}[!h]
    \centering
    \small
    \caption{Experimental results on various datasets for general CF models with and without LLM knowledge transfer}
    \label{tab:results_general}
    \begin{tabular}{cccccccccccccc}
        \toprule
        \textbf{Dataset} & \textbf{CF model} & \multicolumn{3}{c}{\textbf{Recall@10}} & \multicolumn{3}{c}{\textbf{NDCG@10}} & \multicolumn{3}{c}{\textbf{Hits@10}} \\ 
        
        & & Base & LLM-KT & Impr. & Base & LLM-KT & Impr. & Base & LLM-KT & Impr. \\ 
        \midrule
        \multirow{3}{*}{CDs} & NeuMF & 0.1511 & \textbf{0.1579} & 4.50\% & 0.1519 & \textbf{0.1566} & 3.09\% &  0.5855 & \textbf{0.6066} & 3.60\% \\ 
        & SimpleX & 0.1594 & \textbf{0.1708} & 7.15\% &  0.156  & \textbf{0.1669} & 6.99\% &  0.6091 & \textbf{0.6262} & 2.81\% \\ 
        & MultVAE & 0.1451 & \textbf{0.1737} & 19.71\% & 0.1428 & \textbf{0.1736} & 21.57\% & 0.5790 & \textbf{0.6368} & 9.98\% \\ 
        \hline
        \multirow{3}{*}{ML-1M} & NeuMF & {0.098} & \textbf{0.1088} & 11.02\% & {0.18} & \textbf{0.1969} & 9.39\% & {0.7035} & \textbf{0.7325} & 4.12\% \\ 
        & SimpleX & 0.0935 & \textbf{0.108} & 15.51\% & 0.1838 & \textbf{0.2003} & 8.98\% & 0.6899 & \textbf{0.7313} & 6.00\% \\ 
        & MultVAE & 0.1297 & \textbf{0.1352} & 4.24\% & 0.1925 & \textbf{0.1981} & 2.91\% & 0.7281 & \textbf{0.7311} & 0.41\% \\ 
        \bottomrule
    \end{tabular}
    
\end{table*}

\begin{table}[h]
    \centering
    \small
    \caption{AUC-ROC of context-aware models}
    \label{tab:results_kar}
    \begin{tabularx}{\columnwidth}{XXXXXXXXXX}
        \toprule
        \textbf{Dataset} & \textbf{CF model} & \textbf{Base} & \textbf{KAR} & \textbf{LLM-KT} \\
        \midrule
        \multirow{3}{*}{CDs} 
        & DCN & 0.8214 & 0.8204 & \textbf{0.8273} \\
        & DeepFM & 0.8427 & \textbf{0.8477} & 0.8463 \\
        \midrule
        \multirow{3}{*}{ML-1M} 
        & DCN & 0.7753 & 0.7755 & \textbf{0.7889} \\ 
        & DeepFM & 0.7934 & 0.7983 & \textbf{0.8175} \\ 
        \bottomrule
    \end{tabularx}
    
\end{table}

When presenting the results in tables, we use the following notation: mark ``Base.'' is the baseline CF model, ``LLM-KT'' stands for the proposed training of the corresponding baseline with knowledge transfer, and ``KAR'' stands for the baseline model enhanced by KAR.

We tested the proposed method on a reranking task for general CF models. We assessed performance by using ranking metrics such as NDCG@K, Hits@K, and Recall@K. Table \ref{tab:results_general} contains the main results for different baselines. Here, the proposed method consistently enhances the performance of all CF models across considered scenarios. 



We selected the conventional click-through-rate (CTR) prediction task~\cite{shirokikh2024neural} for context-aware models, which was evaluated using the AUC-ROC metric. The experimental results for the proposed method and KAR are shown in Table~\ref{tab:results_kar}. Our method demonstrates consistent performance with KAR. The proposed pretext task of internally reconstructing features proves competitive by explicitly providing them as inputs.


Thus, the proposed knowledge transfer method performs comparably to existing ones but is more versatile, as it can be generalized to any CF model that does not support input features.

\section{Conclusion}

In this work, we present LLM-KT, an experimental framework\footnote{\url{https://github.com/a250/LLMRecSys_with_KnowledgeDistilation/tree/distil_framework}} that enables efficient knowledge transfer from LLMs to CF models. The demonstration video is available here\footnote{\url{https://youtu.be/eVF9EF_oGFw}}.

Leveraging the RecBole platform, LLM-KT seamlessly integrates into diverse applications and existing systems, benefitting from RecBole's comprehensive suite of algorithms, metrics, and methods. This adaptability allows it to support various CF models without requiring architectural modifications, making it suitable for various recommendation tasks. With flexible configuration options, LLM-KT empowers researchers and developers to incorporate LLM-generated features easily, pre-training CF models to harness these embeddings for enhanced performance. 


Our experiments on the MovieLens and Amazon datasets demonstrated that LLM-KT significantly improves the performance of the CF model in general and context-aware scenarios. Notably, the framework is competitive with state-of-the-art approaches such as KAR while offering broader applicability. These results validate the framework's potential for extending the capabilities of CF models through efficient LLM knowledge transfer. Future work will explore alternative architectures, focusing on sequential recommendations, and expand to other domains and datasets.


\bibliographystyle{IEEEtran}
\bibliography{Main}

\begin{thebibliography}{10}
\providecommand{\url}[1]{#1}
\csname url@samestyle\endcsname
\providecommand{\newblock}{\relax}
\providecommand{\bibinfo}[2]{#2}
\providecommand{\BIBentrySTDinterwordspacing}{\spaceskip=0pt\relax}
\providecommand{\BIBentryALTinterwordstretchfactor}{4}
\providecommand{\BIBentryALTinterwordspacing}{\spaceskip=\fontdimen2\font plus
\BIBentryALTinterwordstretchfactor\fontdimen3\font minus
  \fontdimen4\font\relax}
\providecommand{\BIBforeignlanguage}[2]{{%
\expandafter\ifx\csname l@#1\endcsname\relax
\typeout{** WARNING: IEEEtran.bst: No hyphenation pattern has been}%
\typeout{** loaded for the language `#1'. Using the pattern for}%
\typeout{** the default language instead.}%
\else
\language=\csname l@#1\endcsname
\fi
#2}}
\providecommand{\BIBdecl}{\relax}
\BIBdecl

\bibitem{koren2021advances}
Y.~Koren, S.~Rendle, and R.~Bell, ``Advances in collaborative filtering,''
  \emph{Recommender systems handbook}, pp. 91--142, 2021.

\bibitem{shevchenko2024variability}
V.~Shevchenko, N.~Belousov, A.~Vasilev, V.~Zholobov, A.~Sosedka, N.~Semenova,
  A.~Volodkevich, A.~Savchenko, and A.~Zaytsev, ``From variability to
  stability: Advancing {RecSys} benchmarking practices,'' in \emph{Proceedings
  of the 30th ACM SIGKDD Conference on Knowledge Discovery and Data Mining},
  2024, pp. 5701--5712.

\bibitem{kiselev2022exploration}
D.~Kiselev and I.~Makarov, ``Exploration in sequential recommender systems via
  graph representations,'' \emph{IEEE Access}, vol.~10, pp. 123\,614--123\,621,
  2022.

\bibitem{kumar2019predicting}
S.~Kumar, X.~Zhang, and J.~Leskovec, ``Predicting dynamic embedding trajectory
  in temporal interaction networks,'' in \emph{Proceedings of the 25th ACM
  SIGKDD International Conference on Knowledge Discovery \& Data Mining}, 2019,
  pp. 1269--1278.

\bibitem{severin2023ti}
N.~Severin, A.~Savchenko, D.~Kiselev, M.~Ivanova, I.~Kireev, and I.~Makarov,
  ``{Ti-DC-GNN}: Incorporating time-interval dual graphs for recommender
  systems,'' in \emph{Proceedings of the 17th ACM Conference on Recommender
  Systems}, 2023, pp. 919--925.

\bibitem{wang2023enhancing}
Y.~Wang and et~al., ``Enhancing recommender systems with large language model
  reasoning graphs,'' \emph{arXiv preprint arXiv:2308.10835}, 2023.

\bibitem{wu2023leveraging}
J.~Wu, Q.~Liu, H.~Hu, W.~Fan, S.~Liu, Q.~Li, X.-M. Wu, and K.~Tang,
  ``Leveraging large language models ({LLMs}) to empower training-free dataset
  condensation for content-based recommendation,'' \emph{arXiv preprint
  arXiv:2310.09874}, 2023.

\bibitem{wang2023recmind}
Y.~Wang, Z.~Jiang, Z.~Chen, F.~Yang, Y.~Zhou, E.~Cho, X.~Fan, X.~Huang, Y.~Lu,
  and Y.~Yang, ``Recmind: Large language model powered agent for
  recommendation,'' \emph{arXiv preprint arXiv:2308.14296}, 2023.

\bibitem{sun2024large}
Z.~Sun, Z.~Si, X.~Zang, K.~Zheng, Y.~Song, X.~Zhang, and J.~Xu, ``Large
  language models enhanced collaborative filtering,'' \emph{arXiv preprint
  arXiv:2403.17688}, 2024.

\bibitem{xi2023towards}
Y.~Xi and et~al., ``Towards open-world recommendation with knowledge
  augmentation from large language models,'' \emph{arXiv preprint
  arXiv:2306.10933}, 2023.

\bibitem{shu2023rah}
Y.~Shu, H.~Gu, P.~Zhang, H.~Zhang, T.~Lu, D.~Li, and N.~Gu, ``Rah!
  recsys-assistant-human: A human-central recommendation framework with large
  language models,'' \emph{arXiv preprint arXiv:2308.09904}, 2023.

\bibitem{zhang2023agentcf}
J.~Zhang and et~al., ``{AgentCF}: Collaborative learning with autonomous
  language agents for recommender systems,'' \emph{arXiv preprint
  arXiv:2310.09233}, 2023.

\bibitem{mcinnes2018umap}
L.~McInnes, J.~Healy, and J.~Melville, ``Umap: Uniform manifold approximation
  and projection for dimension reduction,'' \emph{arXiv preprint
  arXiv:1802.03426}, 2018.

\bibitem{mackiewicz1993principal}
A.~Ma{\'c}kiewicz and W.~Ratajczak, ``Principal components analysis (pca),''
  \emph{Computers \& Geosciences}, vol.~19, no.~3, pp. 303--342, 1993.

\bibitem{van2008visualizing}
L.~Van~der Maaten and G.~Hinton, ``Visualizing data using {t-SNE},''
  \emph{Journal of machine learning research}, vol.~9, no.~11, 2008.

\bibitem{zhao2021recbole}
W.~X. Zhao and et~al., ``{RecBole}: Towards a unified, comprehensive and
  efficient framework for recommendation algorithms,'' in \emph{Proceedings of
  the 30th ACM International Conference on Information \& Knowledge
  Management}, 2021, pp. 4653--4664.

\bibitem{covington2016deep}
P.~Covington, J.~Adams, and E.~Sargin, ``Deep neural networks for youtube
  recommendations,'' in \emph{Proceedings of the 10th ACM conference on
  Recommender Systems}, 2016, pp. 191--198.

\bibitem{ji2023critical}
Y.~Ji, A.~Sun, J.~Zhang, and C.~Li, ``A critical study on data leakage in
  recommender system offline evaluation,'' \emph{ACM Transactions on
  Information Systems}, vol.~41, no.~3, pp. 1--27, 2023.

\bibitem{he2017neural}
X.~He, L.~Liao, H.~Zhang, L.~Nie, X.~Hu, and T.-S. Chua, ``Neural collaborative
  filtering,'' in \emph{Proceedings of the 26th international conference on
  world wide web}, 2017, pp. 173--182.

\bibitem{mao2021simplex}
K.~Mao and et~al., ``Simplex: A simple and strong baseline for collaborative
  filtering,'' in \emph{Proceedings of the 30th ACM International Conference on
  Information \& Knowledge Management}, 2021, pp. 1243--1252.

\bibitem{liang2018variational}
D.~Liang, R.~G. Krishnan, M.~D. Hoffman, and T.~Jebara, ``Variational
  autoencoders for collaborative filtering,'' in \emph{Proceedings of the 2018
  world wide web conference}, 2018, pp. 689--698.

\bibitem{wang2017deep}
R.~Wang, B.~Fu, G.~Fu, and M.~Wang, ``Deep \& cross network for ad click
  predictions,'' in \emph{Proceedings of the ADKDD'17}, 2017, pp. 1--7.

\bibitem{guo2017deepfm}
H.~Guo, R.~Tang, Y.~Ye, Z.~Li, and X.~He, ``Deepfm: a factorization-machine
  based neural network for ctr prediction,'' \emph{arXiv preprint
  arXiv:1703.04247}, 2017.

\bibitem{shirokikh2024neural}
M.~Shirokikh, I.~Shenbin, A.~Alekseev, A.~Volodkevich, A.~Vasilev, A.~V.
  Savchenko, and S.~Nikolenko, ``Neural click models for recommender systems,''
  in \emph{Proceedings of the 47th International ACM SIGIR Conference on
  Research and Development in Information Retrieval}, 2024, pp. 2553--2558.

\end{thebibliography}

\end{document}